\documentclass[pdflatex,sn-nature]{sn-jnl}

\usepackage{graphicx}%
\usepackage{multirow}%
\usepackage{amsmath,amssymb,amsfonts}%
\usepackage{amsthm}%
\usepackage{mathrsfs}%
\usepackage[title]{appendix}%
\usepackage{xcolor}%
\usepackage{textcomp}%
\usepackage{manyfoot}%
\usepackage{booktabs}%
\usepackage{algorithm}%
\usepackage{algorithmicx}%
\usepackage{algpseudocode}%
\usepackage{listings}%
\usepackage{subcaption}
\usepackage{braket}
\usepackage{bm}

\newcommand{\vR}{\mathbf{R}}

\usepackage{lineno}
\usepackage[normalem]{ulem}

\theoremstyle{thmstyleone}%
\theoremstyle{thmstyletwo}%

\theoremstyle{thmstylethree}%

\definecolor{mediumgreen}{RGB}{0,170,0}

\begin{document}

\title[Article Title]{Molecular Attoscope: Pulse Shape Spectroscopy of Electronic Coherence}


\author[1]{\fnm{Loc Thi-Hoang} \sur{Ngo}}

\author[1]{\fnm{Julia} \sur{Codere}}

\author[1]{\fnm{Javin} \sur{Ohara}}
\author[2]{\fnm{Brian} \sur{Kaufman}}
\author[1]{\fnm{Martin G} \sur{Cohen}}
\author[1]{\fnm{Scott} \sur{Liguori}}
\author[3]{\fnm{Varun} \sur{Makhija}}
\author[4]{\fnm{Tam{\'a}s} \sur{Rozgonyi}}
\author[5]{\fnm{Philipp} \sur{Marquetand}}
\author[2]{\fnm{Matthew} \sur{Bain}}
\author[6]{\fnm{Brett J} \sur{Pearson}}
\author[2,7]{\fnm{Ruaridh} \sur{Forbes}}
\author*[1]{\fnm{Thomas} \sur{Weinacht}}\email{thomas.weinacht@stonybrook.edu}

\affil[1]{\orgdiv{Department of Physics and Astronomy}, \orgname{Stony Brook University}, \orgaddress{\city{Stony Brook}, \state{NY} \postcode{11794-3800}, \country{USA}}}
\affil[2]{\orgdiv{Linac Coherent Light Source}, \orgname{SLAC National Accelerator Laboratory}, \orgaddress{\city{Menlo Park}, \state{CA} \postcode{94025}, \country{USA}}}
\affil[3]{\orgdiv{Department of Chemistry and Physics},
\orgname{University of Mary Washington},
\orgaddress{\city{Fredericksburg}, \state{VA} \postcode{22401}, \country{USA}}}
\affil[4]{\orgdiv{HUN-REN Wigner Research Centre for Physics}, \orgaddress{\city{H-1525 Budapest}, \country{Hungary}}}
\affil[5]{\orgdiv{Quastify GmbH}, \postcode{69124} \orgaddress{\city{Heidelberg}, \country{Germany}}}
\affil[6]{\orgdiv{Department of Physics and Astronomy}, \orgname{Dickinson College}, \orgaddress{\city{Carlisle}, \state{PA} \postcode{17013}, \country{USA}}}
\affil[7]{\orgdiv{Department of Chemistry}, \orgname{University of California, Davis}, \orgaddress{\city{Davis}, \state{CA} \postcode{95616}, \country{USA}}}

\maketitle
\section*{Abstract}
Tracking the coupled motion of electrons and nuclei on their intrinsic timescales is essential to understanding and controlling photochemical transformations. While attosecond techniques have provided unprecedented insight into electronic dynamics, they have largely been restricted to ionic systems, with nuclear motion often neglected or indirectly inferred. Here, we demonstrate a molecular attoscope, which uses shaped laser pulses in the deep ultraviolet to perform coherent measurements of electronic and nuclear dynamics in an entangled wave packet in neutral benzene. This enables us to trace both the 856-attosecond-period electronic motion and the 36-femtosecond-period nuclear motion. We observe electronic coherence persisting over hundreds of optical cycles, modulated by nuclear dynamics. \textcolor{black}{Our interferometric approach is general, and lays the groundwork for coherent  measurements capable of visualizing the evolution of the coupled electronic-nuclear wave function in real time.}

\section*{Introduction}\label{sec1}

The coupled motion of electrons and nuclei lies at the heart of ultrafast science, underpinning our understanding of fundamental processes such as energy transfer and light harvesting \cite{ponseca2017ultrafast,worner2017charge,Scholes2011SolarHarvesting,Yeh2000ElectronLocalization,Maiuri2018FMOCoherence}. The field of femtochemistry allows one to follow molecular transformation in real time, from reactants to products, through the transition state with femtosecond laser pulses that capture the quantum dynamics of molecules \cite{zewail2000femtochemistry,Blanchet1999VibronicDynamics,Gessner2006MultidimensionalImaging,NugentGlandorf2001Br2Dissociation,Dantus1990MolecularVibration}. Attochemistry now aims to push time resolution to the attosecond regime, allowing direct tracking of electronic motion alongside nuclear dynamics\cite{palacios2020quantum,lepine2014attosecond,corkum2007attosecond,goulielmakis2010real,calegari2014ultrafast,nisoli2017attosecond,okino2015direct,matselyukh2022decoherence,kowalewski2015catching,Vacher2017ElectronDynamics,Despre2018ChargeMigration}.   

Although significant progress has been made in tracking electronic processes such as charge migration and transfer in molecules \cite{PhysRevLett.126.133002,matselyukh2022decoherence,Kraus2015,Schwickert2022}, capturing fully coupled electron-nuclear dynamics remains a major challenge. This is partially because attosecond light sources are generally confined to the extreme/vacuum ultraviolet (XUV/VUV) region of the spectrum, where the photon energy is above the ionization potential and couples directly to cationic states of the molecule, limiting the chemical and biological relevance. Moreover, generating attosecond XUV/VUV pump-probe pulses with sufficient energy to yield measurable ionization signals remains technically demanding \cite{Guo2024,kretschmar2024compact}. This has motivated the development of hybrid approaches where one pulse (either the pump or probe) is an attosecond XUV/VUV pulse, while the other is a strong-field, near-infrared pulse. The infrared pulse is easier to generate with sufficient pulse energy to enable strong coupling to the material system of interest \cite{calegari2014ultrafast,matselyukh2022decoherence,okino2015direct,eckle2008attosecond,Belshaw2012ChargeMigration}. While such hybrid approaches have produced fascinating views of attosecond dynamics, it can be difficult to control the quantum dynamics of interest or the details of the final state. They can also be challenging to interpret because it is non-trivial to model the strong-field light-matter interaction. As we discuss below, it is possible to resolve attosecond dynamics in neutral molecules using deep ultraviolet (DUV) femtosecond pulses if one makes use of pulse shaping to control the pulses with attosecond precision.

In this work, we shape laser pulses in the DUV to drive and follow the combined motion of electrons and nuclei in valence states of neutral  molecules. \textcolor{black}{We focus on benzene in this text because it provides a beautiful illustration of how the entanglement between electronic and nuclear degrees of freedom leads to decoherence, but can also show revivals if the dimensionality of the nuclear motion is limited. While previous time-resolved measurements on the S$_1$ ($^1B_{2u}$) state of benzene found evidence of internal conversion (IC) and intersystem crossing (ISC) when exciting at $243$ nm, calculations estimate a barrier of about 3000 $\text{cm}^{-1}$ \cite{parker2009ultrafast}. An initially excited wave packet would have to overcome this barrier to reach points on the excited state surface where non-adiabatic coupling and spin-orbit coupling are sufficient to drive efficient IC and ISC respectively. This means that excitation below the barrier at wavelengths of about 255 nm leads to a long-lived, excited-state wave packet, which facilitates our study.  Furthermore, the interpretation of the measurements is straightforward and provides a framework through which we can describe our experimental method and analytical approach.}

We build on previous work that developed techniques to measure electronic coherences and the formalism to describe them \cite{PhysRevA.110.033118, PhysRevLett.131.263202, mougol2024direct,palacios2014molecular, gonzalez2020quantum}. Our approach enables simultaneous tracking of attosecond-scale electronic motion and femtosecond-scale vibrational motion, offering a clear view of their entangled evolution \cite{PhysRevLett.128.043201,Vrakking2021AttosecondControl}. Although the pulses themselves are not attosecond in duration, our pulse-shaping technique provides sub-$10$ attosecond control over the relative timing between pulses (\textcolor{black}{comparable to actively stabilized interferometric attosecond delay lines \cite{Koll:22}} - see the Supplementary Methods for more details about how we calculate timing stability), \textcolor{black}{which allows us to precisely measure the interference of multiple ionization pathways. Working with femtosecond pulses in the deep UV, as opposed to attosecond pulses in the VUV or XUV, allows us to access neutral excited states for broad photochemical relevance.  Previous experiments have made use of interference to separately reconstruct electronic wavepackets in atomic systems \cite{blanchet1997temporal, feist2011attosecond} and nuclear wavepackets in small molecular systems \cite{engel1994two, blanchet1998temporal, ohmori2006real, mudrich2008quantum} with femtosecond time resolution. Here we access both the electronic and nuclear motion via interfering ionization pathways with attosecond resolution, similar to proposals for H$_2$ by Palacios et al. and Gonz\'alez-Castrillo  et al. \cite{palacios2014molecular, gonzalez2020quantum}. Our measurements differ slightly from their proposal in that we use shaped femtosecond pulses in the deep UV (rather than attosecond pulses in the VUV/XUV) allowing us to excite neutral states of larger molecules  and exert some control over the initially prepared wave packet.} 
\section*{Results}
\subsection*{Time-Resolved Measurements of Quantum Wavefunctions}\label{sec2}

In some sense, the ultimate aim of all time-resolved measurements is to completely characterize the full, time-dependent quantum mechanical wave function \cite{morrigan2023ultrafast, baumgartner2022ultrafast, matselyukh2022decoherence}. For a polyatomic molecule this wavefunction can be written, using a Born-Huang expansion \cite{born1996dynamical}, as a superposition of energy eigenstates in terms of the electronic coordinates $\mathbf{r}_i$, nuclear coordinates $\mathbf{R}_j$, and time, t, as:
\begin{equation}
\Psi(\mathbf{r}_i,\mathbf{R}_j,t)=\sum_n a_{n}\chi_n(\mathbf{R}_j,t)\psi_{n}(\mathbf{r}_i;\mathbf{R}_j)e^{-i \omega_nt},
\label{eq:Born-Huang_expansion}
\end{equation}
where $a_n$ is the complex coefficient representing the amplitude of the $n^{\textrm{th}}$ electronic state in the superposition, $\chi_n(\mathbf{R}_j,t)$ is a normalized vibrational wave function on the $n^{\textrm{th}}$ electronic state, $\psi_{n}(\mathbf{r}_i;\mathbf{R}_j)$ is the $n^{\textrm{th}}$ electronic eigenstate of the system, and $\omega_n=E_n/\hbar$ is the frequency at which the relative phase of each eigenstate (with energy $E_n$) evolves.  The wave function in Equation~\ref{eq:Born-Huang_expansion} evolves on multiple timescales, with the nuclear \textcolor{black}{wave packet dynamics described by $\chi_n(\mathbf{R}_j,t)$ typically occurring on femtosecond timescales, while the dynamics of the electronic wavepacket} driven by the relative phase between electronic eigenstates evolve on attosecond timescales.

\begin{figure}[!htb]
\centering
\includegraphics[width=1\linewidth]{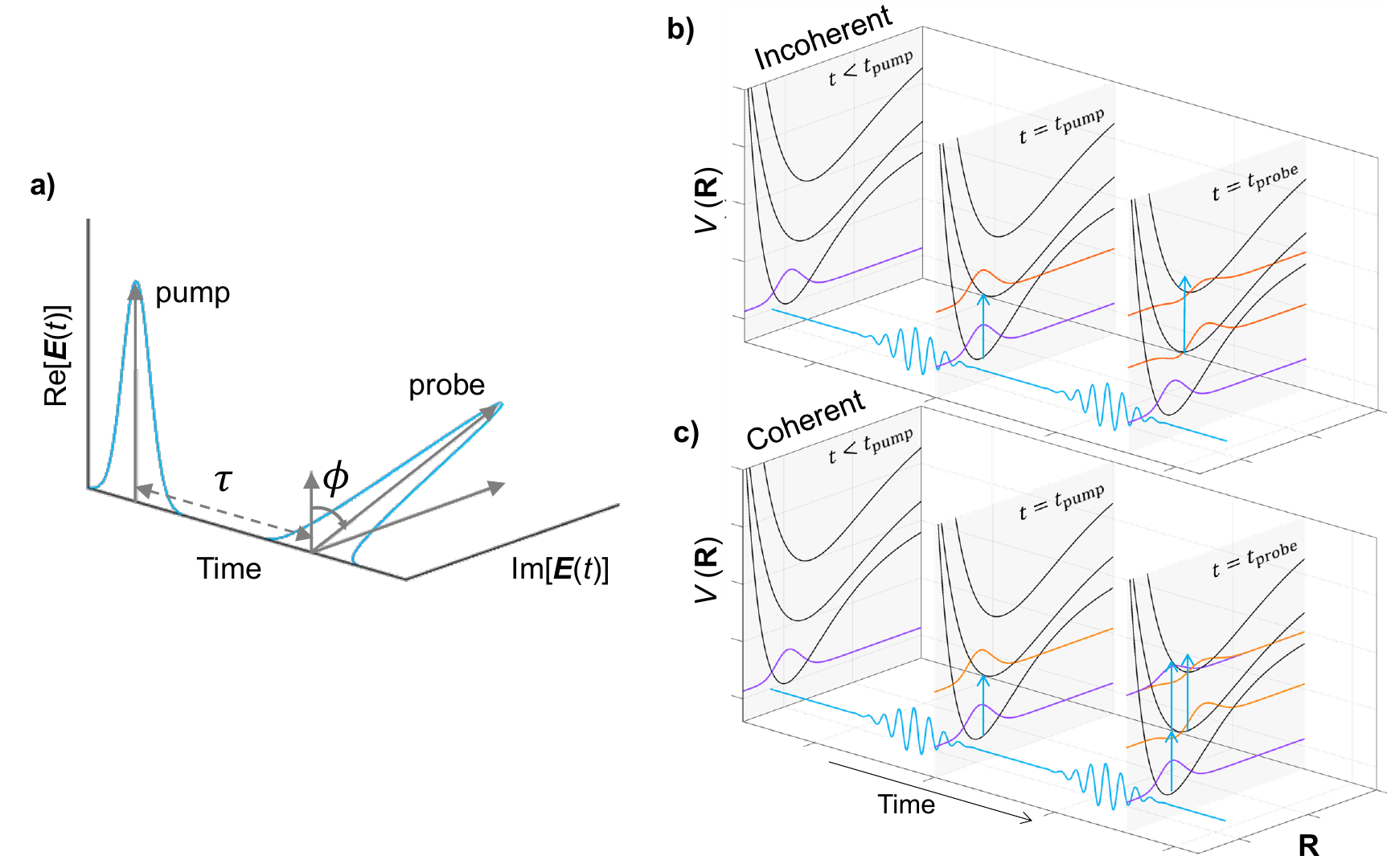}
\caption{\textcolor{black}{\textbf{Illustration of our molecular attoscope for measurements of electronic coherence.}   \textbf{(a)}: Pump-probe pulse sequence with independent control over delay $\tau$ and phase $\phi$ (only the pulse envelopes are shown). Here the relative phase is represented by the angle $\phi$ between the pulses in the complex plane. \textbf{(b)}: Potential energy curves for the ground, excited, and ionic states together with the pulse sequence and vibrational wave functions for a traditional incoherent pump-probe measurement scheme that allows access to vibrational dynamics on the intermediate state of the photoexcited molecule. \textbf{(c)}: Same as panel (b) for our coherent approach measuring both electronic and vibrational dynamics via interfering ionization pathways.} }
\label{fig:scheme}
\end{figure}

Many time-resolved studies in femtochemistry investigate the dynamics of $\chi_n(\mathbf{R}_j,t)$ in~Equation~\ref{eq:Born-Huang_expansion} by tracking the nuclear wave function in the intermediate electronic state $\psi_{n}(\mathbf{r}_i;\mathbf{R}_j)$ using ionization-based probing schemes \cite{stolow2004femtosecond,gerber1991Na2,Blanchet1999VibronicDynamics}. Figure~\ref{fig:scheme}(b) illustrates this approach, where a pump pulse establishes a time-dependent nuclear wave function in an excited electronic state. 
A subsequent probe pulse ionizes the system in a way that is sensitive to the shape and/or position of the evolving wave function, yielding a time-dependent measurement of $\chi_n(\mathbf{R}_j,t)$ (with, for example, $n=2$ for excited-state dynamics). While this approach tracks nuclear dynamics on the excited state, it does not, in general, allow one to measure the electronic coherence (i.e., electron dynamics) established between the eigenstates by the pump pulse, regardless of the laser pulse duration. To be sensitive to the attosecond electron dynamics, one needs to simultaneously probe all of the relevant electronic states in the superposition represented by~Equation~\ref{eq:Born-Huang_expansion} (including the ground state). This generally requires attosecond pulses with bandwidths spanning the energy separation between the electronic states.

Our approach makes use of a pump pulse to create a superposition of electronic states, and a probe pulse to ionize this superposition, \textcolor{black}{similar to the proposal of Gonz\'alez-Castrillo et al. \cite{gonzalez2020quantum}}.  Figure~\ref{fig:scheme} illustrates the pump-probe sequence (panel (a)), and contrasts traditional pump-probe measurements with our interferometric approach (panels (b) and (c), respectively).  In particular, \textcolor{black}{panel (a) shows the pump-probe sequence, with independent control of both the relative phase $\phi$ - represented by the angle between the pulses in the complex plane - and the interpulse delay $\tau$}. This control facilitates the coherent measurements shown in panel (c), in which both the ground and excited electronic states are ionized by the probe pulse, giving rise to interference that allows us to measure their relative phase. 

\subsection*{Attosecond Pulse-Shape Spectroscopy}\label{sec3}
\begin{figure}[!htb]
\centering
\includegraphics[width=1\linewidth]{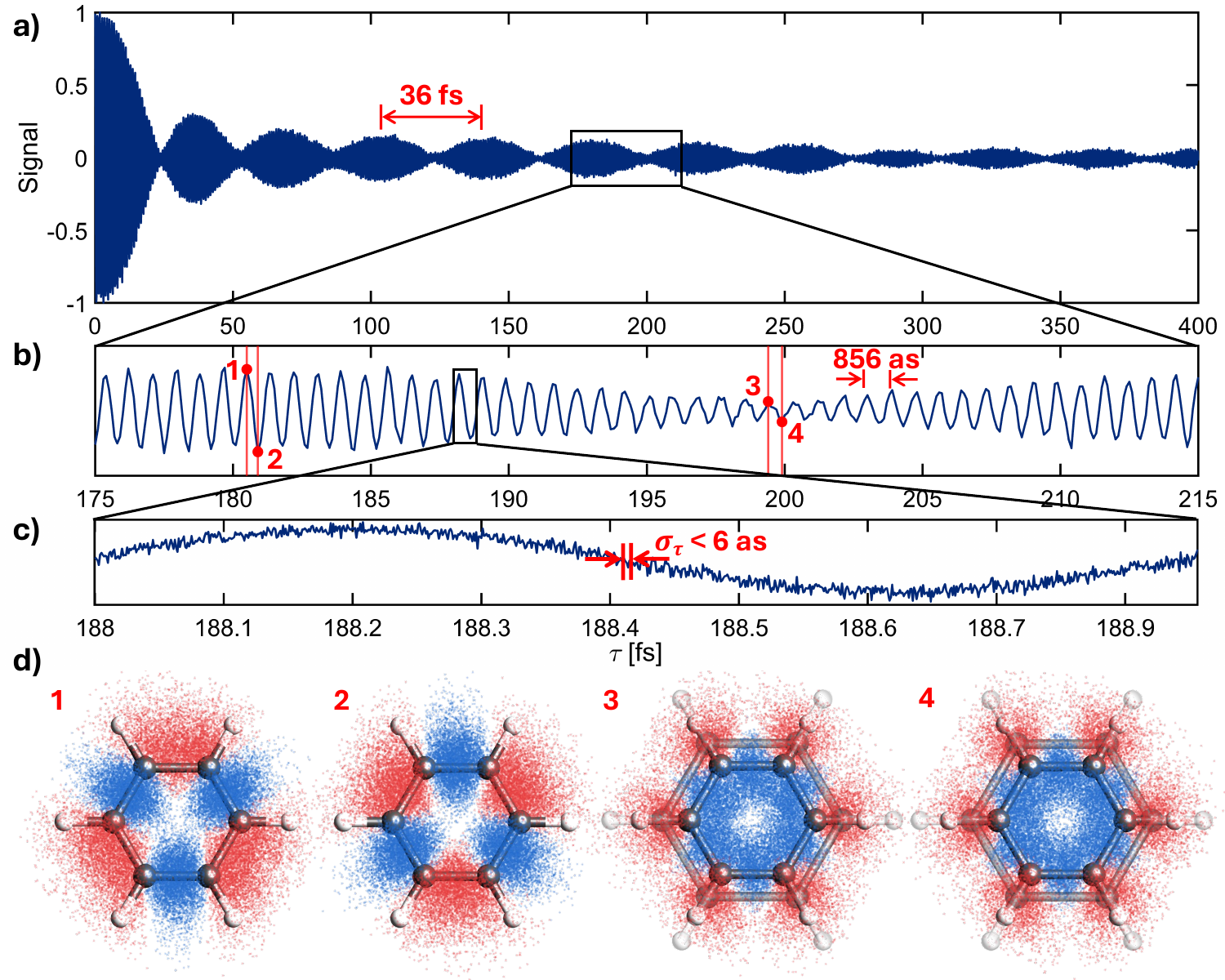}
\caption{\textbf{Molecular Attoscope trace: Differences in ionization yield taken for in-phase and out-of-phase pulses on different timescales.} \textbf{(a)}: On a few hundred femtosecond timescale, the fast oscillations are unresolved but $36$~fs beats can be seen. \textbf{(b)}: Zooming in, faster oscillations under the beats that have an $856$~as period become apparent. \textbf{(c)}: Zooming in even further down to a step size of $1$~as, we can put an upper limit on the phase stability of our apparatus. \textbf{(d)}: Snapshots of the calculated electronic density difference \textcolor{black}{superimposed on evolving nuclear structures representing vibrations in the ground and excited electronic states} (see the Supplementary Movie for detailed calculation) at the specific time delays shown in red in panel (b). The densities of red and blue dots represent positive and negative changes of the electronic density respectively.}
\label{fig:zoom}
\end{figure}

Our experiment consists of measuring the ionization yield for a pair of DUV laser pulses as a function of delay and phase. We make use of a 4--f acousto-optic modulator (AOM)-based pulse shaper to create phase-stable pump and probe pulses \cite{shim2009turn,tian2003femtosecond,pearson2007shaped,codere2024high,catanese2021acousto,codere2025spectral}. More information about the experimental methods are below and in the Supplemental Methods. The experimental data shown in Fig.~\ref{fig:zoom} display a rich interference pattern that contains information on the underlying dynamics in the molecule, \textcolor{black}{similar to the calculations of Gonz\'alez-Castrillo \cite{gonzalez2020quantum}.}

Our pulse shaping capabilities allow us to take the difference between pump-probe measurements when the pulses are in-phase and $\pi$-out-of-phase (see panel (a) of Fig.~\ref{fig:scheme}), highlighting electronic coherences and dynamics, as noted in the calculations below and detailed in the Supplementary Equations. The difference between the ionization yields taken with pulses in-phase and out-of-phase are shown in panels (a)-(c) of Fig.~\ref{fig:zoom}. The panels show increasing levels of temporal zoom, highlighting both the slow nuclear dynamics and rapid electronic timescales, as well as our inherent timing stability. Panel (d) depicts calculated snapshots of the electron density difference at different times (see the Supplementary Movie for details). Note that the electron density changes dramatically between snapshot 1 and 2, in approximately $400$~as, highlighting the need for high precision and stability of the pump-probe delay. Panel (c) shows the achievable temporal resolution by plotting the ionization signal over $1$~fs delay using a $1$~as step-size. From these data, we estimate an upper limit on the uncertainty of the temporal stability of our apparatus of $6$~as (see the Supplementary Methods for uncertainty measurement), \textcolor{black}{ comparable to the stability of other common-path attosecond delay-lines \cite{lee2024few, huppert2015attosecond, borrego2016two,Koll:22}, along with the added benefit of programmable, arbitrary control of the amplitude and phase of each pulse, and a central frequency suited to exciting neutral molecular states.}

Focusing on panels (a) and (b), one can see that there are two oscillatory motions, a beat envelope with a $36$~fs period (due to nuclear dynamics) and $856$~as oscillations (due to electronic dynamics) that are modulated by the beat. The $36$~fs period of the beat envelope allows us to identify nuclear motion along the symmetric breathing mode of the molecule \textcolor{black}{ $\nu_2/\nu_1$ (Herzberg \cite{herzberg1966electronic}/Wilson \cite{wilson1980molecular} assignments}), which has a vibrational frequency of $925$~cm$^{-1}$ (which gives a $36$~fs period). This corresponds to a measurement of the dynamical evolution of $\chi(\mathbf{R}_j,t)$ on the intermediate electronic state. The observed fast oscillation period of $856$~as corresponds to the inverse of the energy separation (the vertical energy separation at the Franck-Condon geometry) between the $^1A_{1g}$ ground and $^1B_{2u}$ excited electronic states in benzene: $(\omega_2-\omega_1)/2\pi =1172$~THz, \textcolor{black}{electronically forbidden because of symmetry, but allowed via Herzberg-Teller coupling facilitated by displacement along the $\nu_{18}/\nu_6$ (Herzberg/Wilson assignment) vibration}. This oscillation represents a measurement of the electronic dynamics due to the superposition of the $\psi_{n}(\mathbf{r}_i;\mathbf{R}_j)$ states in~Equation~\ref{eq:Born-Huang_expansion}. In order to illustrate how the nuclear motion affects the electronic oscillations and leads to the observed modulations in the ionization yield, we show snapshots of a molecular movie  in panel (d) of Fig.~\ref{fig:zoom}, where the numbers note the corresponding delay times indicated in panel (b). The movie was created using the known nuclear vibrational mode and a calculated electronic probability density difference between \textcolor{black}{the time-dependent coherent superposition $|g\rangle $ + $|e\rangle$ excited by the pump pulse and the static ground state $|g\rangle $ prior to the arrival of the pump pulse (calculation and movie in the Supplementary Movie).}

The movie illustrates the entangled molecular wavepacket dynamics by showing both the electronic and nuclear motion together. The superposition of ground and excited electronic states leads to oscillations in the electronic density, which are modulated by the motion of the nuclei. The breathing motion of the nuclei can be visualized as the evolution of the nuclear wavepacket in the excited electronic state. In the snapshots of panel (d) of Fig.~\ref{fig:zoom}, there are two sets of nuclei superimposed.  In snapshots 1 and 2, these two sets are overlapped, while in snapshots 3 and 4, they are well-separated.  The two sets represent the average positions of nuclear wavepackets in the ground and excited states.  When the nuclear wavepackets overlap (snapshots 1 and 2), the electronic density difference oscillates with a large amplitude, leading to large modulations in the ionization yield.  But when the nuclear wavepackets do not overlap (snapshots 3 and 4), the electronic density does not change significantly with time, and the ionization modulations are small. This dynamical interpretation of the measurements is supported by calculations described below.

\subsection*{Analytical Framework for Quantum Decomposition}\label{sec4}

\begin{figure}[!htb]
\centering
\includegraphics[width=1\linewidth]{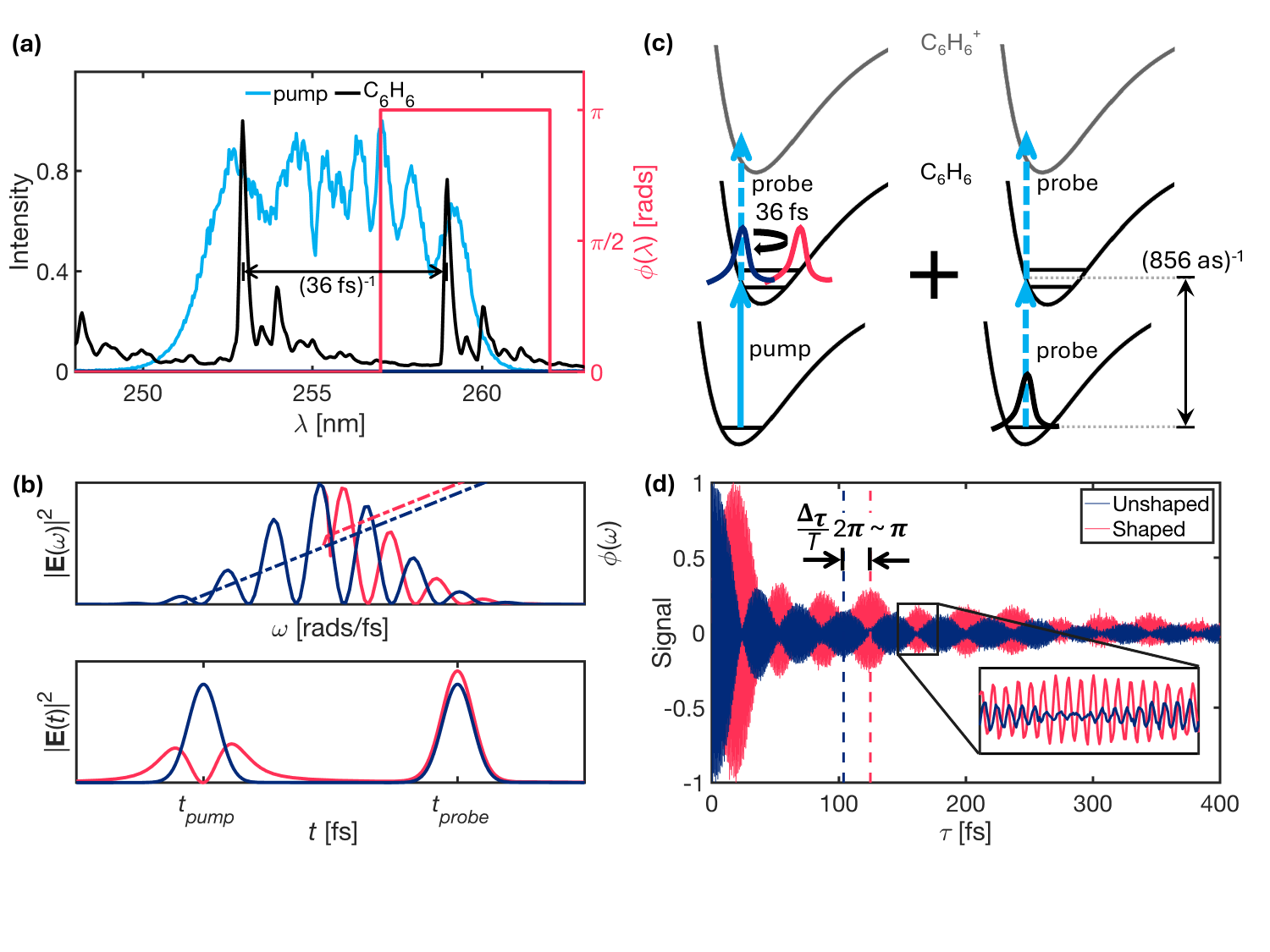}
\caption{\textcolor{black}{\textbf{Molecular interference with shaped pump pulse.} All shaped pulses and signals are shown in red, while all unshaped pulses and signals are are shown in dark blue. \textbf{(a):} Pump-pulse spectrum (light blue) and phase (red) together with the benzene absorption spectrum (black), illustrating a pump pulse shaped to have a $\pi$ phase difference between vibrational eigenstates in the excited-state wave packet. \textbf{(b):} Frequency and time  domain representations of the pump-probe pulse sequence where the pump pulse is shaped to have a spectral phase jump in the middle of the spectrum. The frequency domain plot shows the intensity (solid) and phase (dashed) for the total, combined field of the pump and probe pulses. \textbf{(c):} Two molecular ionization pathways that can interfere.  On the left is single-photon probe ionization of an excited-state wave packet created by a shaped pump pulse such that it starts at the outer turning point (red), which is shifted by half a vibrational period compared to that created by an unshaped pump pulse (dark blue).  On the right is two-photon probe ionization of an unshaped, ground-state wave packet (black). \textbf{(d):} Difference signal for measurements where the pump pulse is shaped to have a spectral phase jump as illustrated in panel (a) (red), together with measurements for an unshaped pump pulse (dark blue).}}
\label{fig:interference}
\end{figure}

\textcolor{black}
{To help facilitate a mathematical interpretation of the results, Fig.~\ref{fig:interference} illustrates a specific implementation of our interferometric approach to wave function measurement. Panel (a) shows the absorption spectrum of benzene together with our laser spectrum. Our laser spectrum covers two primary peaks in the absorption spectrum that correspond to combination bands with one quanta of $\nu_{18}/\nu_6$ (Herzberg/Wilson assignments) and zero quanta of the symmetric breathing mode, $\nu_{2}/\nu_1$, for the peak at $259$~nm, and one quanta of $\nu_{18}/\nu_6$ and one quanta of $\nu_{2}/\nu_1$  for the peak at $253$~nm. We note that the mode $\nu_{18}/\nu_6$ allows the electronic transition via intensity borrowing (Herzberg-Teller coupling), which would otherwise be dark due to symmetry.  Exciting this superposition of vibrational states corresponds to a vibrational wave packet along the symmetric breathing mode in the excited electronic state.  Also shown in this panel is the phase as a function of frequency for the case in which the pump pulse was shaped to introduce a $\pi$ phase shift to frequencies on the long-wavelength side of the spectrum relative to those at shorter wavelengths. This can shift the relative phase of the vibrational eigenstates in the wave packet, allowing for some tailoring of the initial wave packet on the excited state (as shown in the left diagram of panel (c)). Panel (b) plots the intensity of the combined pump-probe pulse sequence in both the frequency and time domains for the shaped pump pulse described above (red), along with those for an unshaped pump-probe sequence (dark blue).  The phase for both cases is also shown in the frequency domain.  Panel (c) illustrates how interference between molecular ionization pathways allows us to measure both electronic and nuclear dynamics together. In this case, the pathways consist of (left) probe-pulse ionization of a shaped excited-state wave packet created by the pump pulse, and (right) probe-pulse ionization of an unshaped ground-state wave packet.  Finally, panel (d) compares the difference signal for the unshaped pump pulse from Fig.~\ref{fig:zoom} to the signal for the phase-shaped pump pulse, which shifts the vibrational wave packet by $\pi$. Further discussion of the shifted data is provided in the Supplementary Discussion.}  

\textcolor{black}{As highlighted in Fig. \ref{fig:interference}, the pulse shaper allows us to shape either pulse so that we can, for example, tailor the initial wave packet and manipulate the phase of interfering ionization pathways.}  In addition to this ability, another key aspect of the pulse shaper is the inherent phase stability of the pump-probe pulse pair. This stability facilitates interferometric measurements of the two ionization pathways shown in panel (b) of Fig.~\ref{fig:interference}, uncovering the motion of both electrons and nuclei. Without this phase stability, we would not be able to resolve the interference in the ionization yield from the two electronic states, and the signal would include only the $36$~fs envelope due to nuclear oscillations (on top of a background arising from ground-state ionization). \textcolor{black}{Finally, control over the relative phase $\phi$ between the pulses allows us to combine pump-probe measurements with different phases \cite{tian2003femtosecond,hamm2011concepts,Fuller20152DES} to highlight specific aspects of the dynamics.  Measurements of the ionization yield as a function of pump-probe delay for different phases are shown in the Supplementary Discussion.}

Motivated by the approach described above, we provide here a simple mathematical formalism that describes the measurements and how they can be directly related to elements of the quantum mechanical wavefunction and electronic density matrix (a detailed derivation is provided in the Supplementary Equations, which includes Ref. \cite{potts1972photoelectron}).  \textcolor{black}{The approach is tailored to the case of benzene, but the derivations are general and can be applied to other molecules.  For example, measurements for acetylene and fluorobenzene are provided in the Supplementary Discussion.}

\textcolor{black}{Absorption of the pump pulse creates a coherent superposition of electronic and vibrational states, which for the case of two electronic states can be written as \cite{weinacht2018time}}
\begin{align}
    \Psi(\mathbf{r}_i,\mathbf{R}_j,t) &= a_0\psi_0(\mathbf{r}_i;\mathbf{R}_j) e^{-i\omega_0t} \chi_{0}(\mathbf{R}_j,t) + a_1\psi_1(\mathbf{r}_i;\mathbf{R}_j) e^{-i\omega_1t} \chi_{1}(\mathbf{R}_j,t)
    \label{eq:two_state}
\end{align}
Absorption of the probe pulse then leads to ionization, and the signal as a function of delay and relative phase between the pump and probe can be written as: 
{\color{black} 
\begin{align}
    S(\tau,\phi) &= |a_0E_0^2Q_{c0}|^2 + |a_1E_0Q_{c1}|^2\\ \nonumber
    &+a_0^*a_1Q^*_{c0}E_0^3Q_{c1}e^{-i\phi}
    e^{-i(\omega_1-\omega_0)\tau}
    \braket{\chi_{0}(\vR_j,\tau) | \chi_{1}(\vR_j,\tau)} + \textup{c.c.}
    \label{eq:big_kahuna}
\end{align}
where $E_0$ represents the electric field amplitude for the pump and probe pulses and 
\begin{equation}    
    Q_{cn} \equiv \braket{ \psi_c(\mathbf{r}_i;\vR_j)  |Q^{(q)}(\mathbf{r}) | \psi_n(\mathbf{r}_i;\vR_j)  } 
\end{equation}
where \textcolor{black}{$\psi_c(\mathbf{r}_i;\vR_j)$ represents the $N-1$ electronic wave function of the cation and} $Q^{(q)}(\mathbf{r})$ is the $\mathbf{r}$  dependence of  $Q^{(q)}(\mathbf{r},t)$, a multiphoton ionization operator, with $q$ being the multiphoton order (either one or two).
}

The difference signal $S(\tau,0)-S(\tau,\pi)$ isolates the off-diagonal elements of the electronic density matrix, while the diagonal terms come from $S(\tau,0)+S(\tau,\pi)$:
\begin{align}
    S(\tau,0)-S(\tau,\pi) &=2Q^*_{c0}Q_{c1}\rho_{01}(\tau) + \textup{c.c.}\\
    S(\tau,0)+S(\tau,\pi) 
    &=2\rho_{00}(\tau)|Q_{c0}|^2+2\rho_{11}(\tau)|Q_{c1}|^2
\end{align}
where 
\begin{align}
    \rho_{mn}(\tau) = a^*_ma_ne^{-i(\omega_n-\omega_m)\tau} \braket{\chi_m(\mathbf{R}_j,\tau) |\chi_n(\mathbf{R}_j,\tau)}
\end{align}
Importantly, while the individual signals $S(\tau,0)$ and $S(\tau,\pi)$ mix diagonal and off-diagonal elements of the density matrix, the difference signal directly measures the off-diagonal elements (electronic coherences), whereas the sum signal directly measures the diagonal elements (populations), and could thus be used to follow non-adiabatic population transfer (internal conversion).

\textcolor{black}{\subsection*{Tests of Molecular Interference}}\label{sec5}

\textcolor{black}{An important aspect of interferometric measurements of molecular coherence is the ability to discriminate between molecular and optical interference in order establish that the measurements are sensitive to the phase between quantum states and not just the spectral intensity of the total applied field (pump+probe) at resonant peaks in the molecular absorption spectrum.  In order to demonstrate that our measurements are a result of interfering molecular ionization pathways, we performed additional experiments where we eliminated optical interference by removing the frequency components that correspond to the primary peaks in the molecular absorption spectrum from the probe pulse. }

\begin{figure}[!htb]
\centering
    \includegraphics[width=0.95\linewidth]{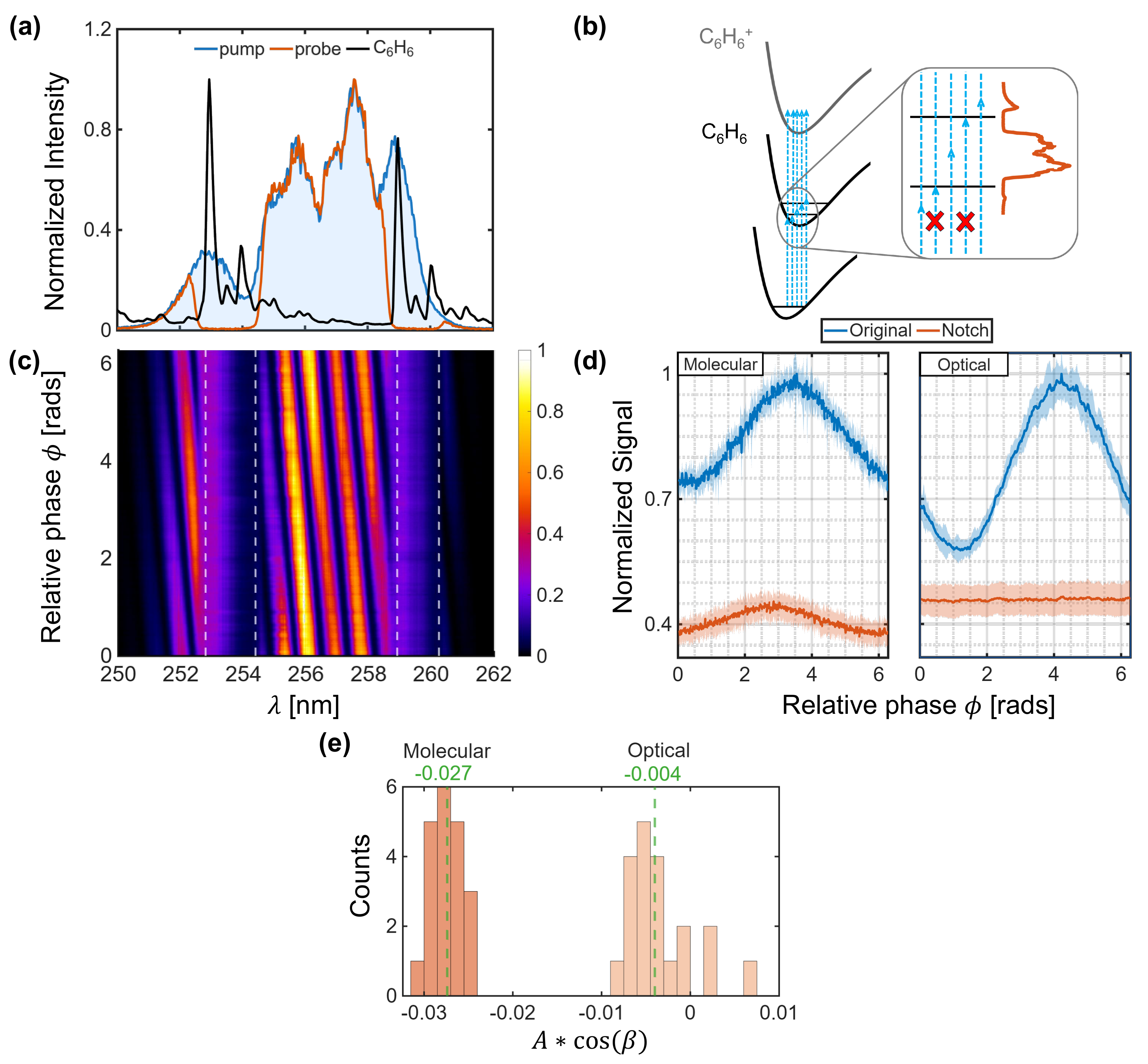}
    \caption{\textcolor{black}{{\textbf{Molecular interference with shaped probe pulse.}} \textbf{(a)}: Benzene absorption spectrum (black), pump-pulse spectrum (blue), and probe-pulse spectrum (orange) with notches at wavelengths corresponding to primary peaks and hot bands in the benzene absorption spectrum. Notch positions: left, $252.8$--$254.4\,\mathrm{nm}$; right, $258.9$--$260.2\,\mathrm{nm}$.  \textbf{(b)}: Ionization pathways from the ground electronic state, with red X's indicating which pathways are eliminated with the notch spectrum. \textbf{(c)}: Optical spectrum of the combined pump and probe pulses as a function of both wavelength and relative phase between pulses. \textbf{(d)}: Averaged normalized molecular (left panel) and optical (right panel) signals as a function of relative phase between pump and probe pulses at a fixed delay ($261.2$~fs) for both unshaped and notched probe spectra. The shaded regions around each curve denote $\pm 1$ standard deviation calculated from all measurements in the corresponding dataset. \textbf{(e)}: Distribution of $A\cos(\beta)$ extracted from 20 independent molecular and optical measurements with a notched probe pulse spectrum. For each measurement, the normalized signal was fitted to cosine waves: Signal($\phi$)=$A\cos(\phi+\beta)+C$. The histograms show the frequency (counts) of the resulting $A\cos(\beta)$ values. The dashed green lines denote the mean of the distributions.}}
    \label{fig:notch}
\end{figure}

\textcolor{black}{As illustrated in Fig. \ref{fig:notch}a, we put two notches in the spectrum of the probe pulse at the positions of the strong benzene absorption peaks, such that there is no optical interference between the pump and probe at these locations (the spectrum of the pump pulse remained unchanged). We then measured the molecular ionization signal using the TOF-MS, as well as the optical spectrum of the combined pulses after the pulse shaper with an optical spectrometer, as a function of the relative phase between pump and probe pulses. Fig.~\ref{fig:notch}b illustrates the modified molecular interference experiment by showing how the probe pulse does not contain frequencies resonant with the ground to excited-state transitions (no optical interference pathways remaining as shown in the zoomed-in region), while still containing near-resonant, two-photon molecular pathways to ionization (i.e., two-photon pathways that are close to single-photon resonance). Fig. \ref{fig:notch}c shows the optical interference data as a function of both wavelength and relative phase between pump and probe.  As expected, the combined optical spectrum shows no interference in the regions where we applied the notches to the probe spectrum (between dashed white lines).}

Fig. \ref{fig:notch}d shows the results of the molecular \textcolor{black}{(left panel) and optical (right panel)} measurements as a function of the relative phase between the pump and the probe, demonstrating that, with the notched probe spectrum, we still see molecular interference, but no optical interference. \textcolor{black}{In the left panel, the molecular ionization signal is shown in blue for the full probe spectrum and in orange for the probe with spectral notches. As expected from Equation~3, interference between pathways produces oscillations with a period of $2\pi$. Although the modulation depth is reduced, the interference persists. In the right panel, the blue and orange curves represent the integral of the benzene absorption spectrum multiplied by the optical spectrum, plotted as a function of the relative phase between the pump and probe pulses for the full and notched probe spectra respectively.} By multiplying the spectrum of the pump and probe pulses by the molecular absorption spectrum, we fully account for optical interference at any features in the absorption spectrum, even small peaks outside the notch regions. This provides a very stringent test of whether optical interference can explain our measurement results.  These measurements demonstrate that even in the absence of optical interference, we see clear interference in the molecular ionization yield, confirming that the phase-dependent yield is a result of interference between molecular ionization pathways and highlighting the capability of the pulse shaper to discriminate between different interfering pathways.
\textcolor{black}{Finally, in Fig.~\ref{fig:notch}e we show histograms of the fitted amplitude times the cosine of the phase for individual scans that are averaged in Fig.~\ref{fig:notch}d. We chose this x-axis for the histogram to highlight the fact that the optical measurements yielded fits with very small amplitudes and widely varying phases (yielding $A\cos(\beta)\sim0$), while the molecular measurements were much more consistent and yielded a value far from zero.  The Supplementary Discussion shows additional tests of molecular interference.} 

\textcolor{black}{\subsection*{Extension to Other Systems}}\label{sec5}

For our excitation wavelengths,  benzene is an example of a molecule with long-lived electronic coherence, driven by single-photon absorption and modulated by vibrational motion along primarily one degree of freedom. We contrast this in the Supplementary Discussion with measurements on fluorobenzene (C$_6$H$_5$F), which shows a more complicated and rapid decay of the electronic coherence as a result of vibrational dynamics along multiple degrees of freedom.
Our approach also allows us to measure coherences between multiple electronic states, with single or multiphoton coherences enabled by measurements with different combinations of phases between the pulses. Specifically, odd photon-order coherences can be isolated in the difference signal, while even photon-order coherences can be found in the sum signal, together with the populations as noted above.  Fourier analysis can be used to separate multiphoton coherences from population evolution.

As an example of a multiphoton coherence, we present measurements of a two-photon coherence in acetylene (C$_2$H$_2$) in the Supplementary Discussion.  
These measurements in acetylene show electronic oscillations at roughly twice the frequency of the benzene and fluorobenzene measurements ($\sim430$ as period instead of $\sim 860$ as). This is highlighted by a Fourier analysis of the measurements shown in Supplementary Fig.~11.  This demonstrates that electronic coherences can be generated at multiple photon orders, thereby expanding the manifold of accessible states. Consequently, the range of electronic states that can contribute to the wave packet is no longer strictly limited by the pulse duration, which typically constrains participation within a given photon order. Further measurements on a series of other molecules have been performed, and a detailed interpretation of multiphoton coherences will be presented in upcoming manuscripts.

Finally, we note that in addition to measuring electronic coherences, our approach also allows us to manipulate them by tailoring the spectral phase of the pump pulse or adding additional pulses to the sequence. For instance, we can shift the phase of the electronic or nuclear motion by tailoring the pump pulse, as shown in Figs. \ref{fig:zoom} and \ref{fig:interference} above. 

\bigskip

\subsection*{Numerical Modeling of Light-Matter Interaction}\label{sec5}

In order to simulate the measurement observable and confirm \textcolor{black}{our interpretation of the dynamics}, we numerically solve the time-dependent Schr\"{o}dinger equation (TDSE) using a simple but complete model described in earlier work \cite{rozgonyi2000application, PhysRevA.110.033118,PhysRevLett.131.263202}. These simulations include two neutral electronic states (the $^1A_{1g}$ ground state and the $^1B_{2u}$ excited state) plus a discretized electronic continuum.  They allow for nuclear dynamics along one degree of freedom (the symmetric breathing mode \textcolor{black}{$\nu_{2}/\nu_1$ in Herzberg/Wilson assignments)}, as well as the light-matter interaction with both the pump and probe laser fields. Input parameters include the potential energy curves, transition frequencies, coupling strengths, and ionization transition dipole moments; measured values are used for all molecular parameters except the ionization dipole moments \cite{herzberg1966electronic,pantos1978extinction}. \textcolor{black}{The simulations include contributions from hot bands ($\nu_2'+\nu_{18}'+\nu_{20}'-\nu_{20}''$ and $\nu_{18}'+\nu_{20}'-\nu_{20}''$ weighted in proportion to their peak heights relative to the main peaks ($\sim$ 0.12) in the absorption spectrum - see Supplementary Fig.~5) and} are performed for the exact pulse sequences used in the experiments, \textcolor{black}{with constant total (pump+probe) pulse energy for all delays,} covering pulse delays between $0$ and $400$~fs, using $100$~as steps and two different relative phases between the pump and probe field ($0$ and $\pi$). The resulting ionization yields were subsequently combined by taking their differences, allowing direct comparison with experimental results. Further details are provided in the Supplementary Discussion.

Figure~\ref{fig:seawulf} shows the results of the TDSE calculations in a format similar to Fig.~\ref{fig:zoom}. 
\begin{figure}[!htb]
\centering
\includegraphics[width=1\linewidth]{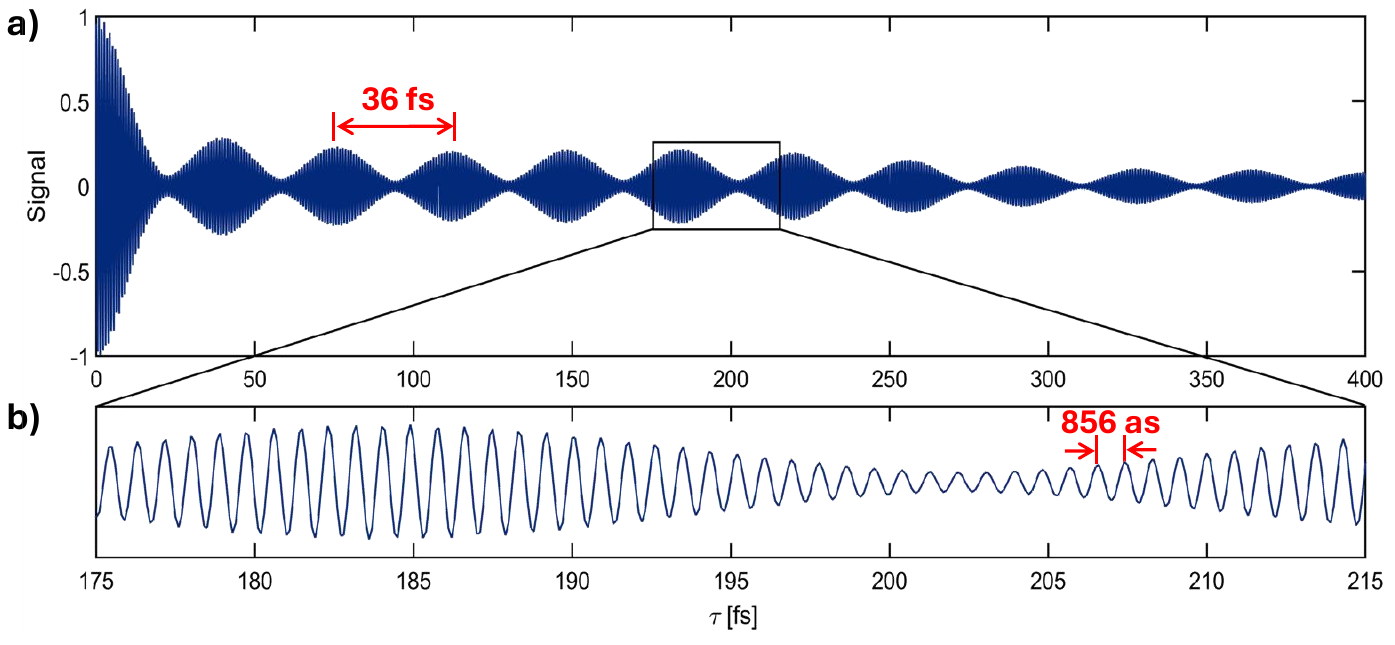}
\caption{\textcolor{black}{\textbf{Simulation of the measurement observable.} Calculated ionization yield as a function of pump-probe delay from solving the TDSE. Solutions for $\pi$ phase difference between pump and probe pulses are subtracted from those for $0$ phase difference, as with the experimental measurements.  All electric field parameters are chosen to be similar to the experimental measurements (pulse durations, fixed pulse energies, delay step size, etc). \textbf{(a)} Calculations from $0$ to $400$~fs delay, using a time step of $100$~as. \textbf{(b)} Zoom in on the calculations for the same delay range as shown in panel (b) of Fig. \ref{fig:zoom}}. 
\label{fig:seawulf}
}
\end{figure}
We note the excellent agreement between theory and experiment, with both displaying rapid, attosecond oscillations with an $856$~as period (highlighted in panel (b)), modulated by a much slower envelope with a period of $36$~fs (as seen in panel (a)). We note that while the measurements and calculations agree on the frequencies of the fast and slow modulations, there is a small phase shift between the measurements and calculations in the low frequency modulations. As detailed in the Supplementary Discussion, this phase shift can be ascribed to small differences in the second-order spectral phase of the pulses in the experiment compared to the calculations. The overall agreement between the TDSE simulations and our measurements helps confirm our earlier interpretation of the measurements.

\textcolor{black}{For clarity, we have focused on coupled electron-nuclear dynamics that involve motion primarily along one normal-mode coordinate; future work aims to extend the approach to coupled dynamics in multiple dimensions.}  This will require calculations capable of incorporating many nuclear degrees of freedom simultaneously, and we have begun to carry out Multiconfigurational Time-Dependent Hartree (MCTDH) simulations \cite{Meyer1990} that make use of fitted ab initio potentials and treat nuclear dynamics along multiple dimensions simultaneously. This, together with our pulse shaping capabilities, will allow us to extend our measurements to larger molecular systems with more complicated nuclear dynamics and more explicit non-adiabatic coupling between electronic and nuclear degrees of freedom.

\textcolor{black}{We have demonstrated a molecular attoscope, which provides a new approach for tracking coupled electronic and nuclear dynamics in neutral molecules using deep-ultraviolet pulse shaping with sub-10 attosecond stability.} Our experiment creates and \textcolor{black}{manipulates} an attosecond-scale electronic coherence modulated by femtosecond nuclear motion in neutral states of benzene, a prototypical aromatic molecule. The ability to coherently probe both degrees of freedom and isolate their contributions via controlled pulse phase manipulation allows us to cleanly access the structure of the full time-dependent vibronic wavefunction. \textcolor{black}{Supported by analytical modeling and numerical simulations that solve the time-dependent Schr\"{o}dinger equation, our results reveal the entangled nature of electronic and vibrational motion and validate an experimental route to \textcolor{black}{interferometric reconstruction of} molecular wave packets \cite{gonzalez2020quantum} by measuring the relative phase between electronic eigenstates in superposition}. This work marks a significant step toward realizing molecular movies that fully capture the ultrafast quantum evolution of complex systems in real time, with implications for understanding light-driven chemistry, biology, and material science.

\section*{Methods}

We use ultrafast pulses in the deep ultraviolet (DUV) to both photoexcite and ionize gas-phase benezene molecules. A ytterbium laser (Light Conversion Pharos) produces $\sim$$250$~fs pulses in the near infrared ($1025$~nm), from which we generate the fourth harmonic in the DUV ($256$~nm) using a pair of second-harmonic-generation crystals. This light propagates through a two-meter-long, stretched, hollow-core fiber (S-HCF) that is filled with argon or krypton gas and undergoes Kerr-induced self-phase-modulation to broaden the spectral content. The bandwidth achieved supports transform-limited pulse durations under $20$~fs (see panel (a) of Fig.~\ref{fig:interference}). The pulses are compressed to $\sim18$~fs,  characterized using frequency resolved optical gating, FROG, (as seen in Supplementary Fig. 1), and shaped using an acousto-optic (AOM)-based pulse shaper, similar to previous work in the visible and near-IR \cite{catanese2021acousto, codere2024high}. The shaper is set-up as a 4--$f$ zero-dispersion stretcher with the AOM placed at the Fourier plane. The electrical signal from an arbitrary waveform generator drives the transducer in the AOM to produce an acoustic pulse, which acts as a spectral mask function. We shape the output laser pulse, $E^{'}(\omega)$, by applying the spectral mask, $M(\omega)$, to the input pulse, $E(\omega)$, such that $E^{'}(\omega)=M(\omega)E(\omega)$. The shaper permits control of the amplitude and phase of the shaped pulse via the mask function $M(\omega)=A(\omega)\exp(i\varphi(\omega))$.

We initially compress the pulses from the S-HCF using a phase mask in the form of a Taylor series expansion:
\begin{equation}
\varphi(\omega) = \frac{\text{GDD}}{2!}(\omega - \omega_0)^2 + \frac{\text{TOD}}{3!}(\omega - \omega_0)^3 + \frac{\text{FOD}}{4!}(\omega - \omega_0)^4
\label{eq:phase mask}
\end{equation}
where the appropriate GDD, TOD, and FOD are applied to minimize the pulse duration. The range of phase that can be applied by the pulse shaper is primarily limited by the acoustic bandwidth of the AOM. This limits us to roughly $\pm 10,000$ fs$^2$ GDD, $\pm 100,000$ fs$^3$ TOD, and $\pm  1.5 \times 10^6$ fs$^4$ FOD. However, the range for each term depends on the values for the previous and consecutive terms. 

We then apply an additional (multiplicative) complex amplitude mask to create a pulse-pair:
 \begin{equation}
M(\omega)=1+\exp \left(i\omega \tau + i\phi \right)
\label{eq:amp mask}
 \end{equation}
where $\tau$ is the delay and $\phi$ is the relative phase between pump and probe pulses. This mask function results in the DUV pulse-pairs used in the pump-probe experiments discussed above. The relative amplitudes, phases, and delays of these pulses are all precisely controllable, and the system achieves sub-$10$ attosecond temporal stability.

We transfer liquid-phase samples to a manifold which is connected to the vacuum chamber. The air in the manifold is removed using the freeze-pump-thaw method before the sample vapor flows to the molecular nozzle. We then measure the total ionization yield of benzene in a time-of-flight mass spectrometer (TOF-MS) as a function of delay between the pump and probe pulses. The experiment is run twice, once with the pulses in-phase ($\phi=0$) and once with the pulses out-of-phase ($\phi=\pi$). Then, the sum or difference of the data is taken. This approach is similar to techniques in two-dimensional spectroscopy, where particular pulse sequences and phase-cycling are used to pick out the molecular signal of interest \cite{tian2003femtosecond, brixner:2D_dots}. We note that our pulse shaper allows us to run pump-probe measurements with a randomized sequence of delays in order to mitigate the impact of slow drifts in the experimental conditions when averaging the results of multiple pump-probe scans. For each pump-probe delay, we averaged 200 laser shots.  A typical pump-probe scan from 0-400 fs delay with 100 as step size for each phase takes about 15 minutes. An average of 5 scans thus takes slightly over an hour.  We found that our laser power and molecular densities were stable over these times.

\backmatter

\bmhead{\textcolor{black}{Data Availability}}
\textcolor{black}{The data supporting the plots in this study are deposited at https://doi.org/10.6084/m9.figshare.31224808 (ref.\cite{data})}.
\bmhead{\textcolor{black}{Code Availability}}
\textcolor{black}{The code used for solving the TDSE and making the movie are available at https://doi.org/10.6084/m9.figshare.31224883}

\bmhead{Acknowledgements}
We gratefully acknowledge Spiridoula Matsika, and Benjamin G. Levine for helpful discussions. 

\bmhead{Funding}
L.T-H.N., J.C., J.O., S.L., and T.W. were funded by the National Science Foundation under award number 2409596. B.J.P. was funded by Dickinson College. B.K. and M.B. were funded by the Linac Coherent Light Source, SLAC National Accelerator Laboratory, which is supported by the U.S. Department of Energy, Office of Science, Office of Basic Energy Sciences, under Contract No. DE-AC02-76SF00515. R.F. was supported by the U.S. Department of Energy (DOE), Office of Science, Office of Basic Energy Sciences, under Award No. DE-SC0026078.  T.R. was funded by the Hungarian National Research, Development and Innovation Fund under grant numbers TKP2021-NVA-04 and NKFIH FK 145967.
\bmhead{Author contributions}
The work was conceptualized by B.J.P., R.F. and T.W.. L.T-H.N., J.C., J.O., B.K.,
M.G.C., S.L., V.M.,  M.B., B.J.P., R.F., and T.W. conducted the experimental measurements, which were analyzed by L.T-H.N. , J.C. and V.M.. L.T-H.N., J.O., B.K. , T.R., and P.M. performed the numerical modeling. The manuscript was written by L.T-H.N., J.C., B.J.P., R.F., and T.W. with input from all authors.
\bmhead{Competing interests}
There are no competing interests to declare.

\section*{Figure Captions}
Figure 1: {\textbf{Illustration of our molecular attoscope for measurements of electronic coherence.}   \textbf{(a)}: Pump-probe pulse sequence with independent control over delay $\tau$ and phase $\phi$ (only the pulse envelopes are shown). Here the relative phase is represented by the angle $\phi$ between the pulses in the complex plane. \textbf{(b)}: Potential energy curves for the ground, excited, and ionic states together with the pulse sequence and vibrational wave functions for a traditional incoherent pump-probe measurement scheme that allows access to vibrational dynamics on the intermediate state of the photoexcited molecule. \textbf{(c)}: Same as panel (b) for our coherent approach measuring both electronic and vibrational dynamics via interfering ionization pathways.}\newline

Figure 2: \textbf{Molecular Attoscope trace: Differences in ionization yield taken for in-phase and out-of-phase pulses on different timescales.} \textbf{(a)}: On a few hundred femtosecond timescale, the fast oscillations are unresolved but $36$~fs beats can be seen. \textbf{(b)}: Zooming in, faster oscillations under the beats that have an $856$~as period become apparent. \textbf{(c)}: Zooming in even further down to a step size of $1$~as, we can put an upper limit on the phase stability of our apparatus. \textbf{(d)}: Snapshots of the calculated electronic density difference \textcolor{black}{superimposed on evolving nuclear structures representing vibrations in the ground and excited electronic states} (see the Supplementary Movie for detailed calculation) at the specific time delays shown in red in panel (b). The densities of red and blue dots represent positive and negative changes of the electronic density respectively.\newline

Figure 3: \textcolor{black}{\textbf{Molecular interference with shaped pump pulse.} All shaped pulses and signals are shown in red, while all unshaped pulses and signals are are shown in dark blue. \textbf{(a):} Pump-pulse spectrum (light blue) and phase (red) together with the benzene absorption spectrum (black), illustrating a pump pulse shaped to have a $\pi$ phase difference between vibrational eigenstates in the excited-state wave packet. \textbf{(b):} Frequency and time  domain representations of the pump-probe pulse sequence where the pump pulse is shaped to have a spectral phase jump in the middle of the spectrum. The frequency domain plot shows the intensity (solid) and phase (dashed) for the total, combined field of the pump and probe pulses. \textbf{(c):} Two molecular ionization pathways that can interfere.  On the left is single-photon probe ionization of an excited-state wave packet created by a shaped pump pulse such that it starts at the outer turning point (red), which is shifted by half a vibrational period compared to that created by an unshaped pump pulse (dark blue).  On the right is two-photon probe ionization of an unshaped, ground-state wave packet (black). \textbf{(d):} Difference signal for measurements where the pump pulse is shaped to have a spectral phase jump as illustrated in panel (a) (red), together with measurements for an unshaped pump pulse (dark blue).}\newline

Figure 4: \textcolor{black}{{\textbf{Molecular interference with shaped probe pulse.}} \textbf{(a)}: Benzene absorption spectrum (black), pump-pulse spectrum (blue), and probe-pulse spectrum (orange) with notches at wavelengths corresponding to primary peaks and hot bands in the benzene absorption spectrum. Notch positions: left, $252.8$--$254.4\,\mathrm{nm}$; right, $258.9$--$260.2\,\mathrm{nm}$.  \textbf{(b)}: Ionization pathways from the ground electronic state, with red X's indicating which pathways are eliminated with the notch spectrum. \textbf{(c)}: Optical spectrum of the combined pump and probe pulses as a function of both wavelength and relative phase between pulses. \textbf{(d)}: Averaged normalized molecular (left panel) and optical (right panel) signals as a function of relative phase between pump and probe pulses at a fixed delay ($261.2$~fs) for both unshaped and notched probe spectra. The shaded regions around each curve denote $\pm 1$ standard deviation calculated from all measurements in the corresponding dataset. \textbf{(e)}: Distribution of $A\cos(\beta)$ extracted from 20 independent molecular and optical measurements with a notched probe pulse spectrum. For each measurement, the normalized signal was fitted to cosine waves: Signal($\phi$)=$A\cos(\phi+\beta)+C$. The histograms show the frequency (counts) of the resulting $A\cos(\beta)$ values. The dashed green lines denote the mean of the distributions.}\newline

Figure 5: \textcolor{black}{\textbf{Simulation of the measurement observable.} Calculated ionization yield as a function of pump-probe delay from solving the TDSE. Solutions for $\pi$ phase difference between pump and probe pulses are subtracted from those for $0$ phase difference, as with the experimental measurements.  All electric field parameters are chosen to be similar to the experimental measurements (pulse durations, fixed pulse energies, delay step size, etc). \textbf{(a)} Calculations from $0$ to $400$~fs delay, using a time step of $100$~as. \textbf{(b)} Zoom in on the calculations for the same delay range as shown in panel (b) of Fig. \ref{fig:zoom}}.


\begin{thebibliography}{10}

\expandafter\ifx\csname url\endcsname\relax
  \def\url#1{\burl{#1}}\fi
\expandafter\ifx\csname urlprefix\endcsname\relax\def\urlprefix{URL }\fi
\providecommand{\bibinfo}[2]{#2}
\providecommand{\eprint}[2][]{\url{#2}}
\providecommand{\doi}[1]{\url{https://doi.org/#1}}
\bibcommenthead

\bibitem{ponseca2017ultrafast}
\bibinfo{author}{Ponseca~Jr, C.~S.}, \bibinfo{author}{Chabera, P.}, \bibinfo{author}{Uhlig, J.}, \bibinfo{author}{Persson, P.} \& \bibinfo{author}{Sundstrom, V.}
\newblock \bibinfo{title}{Ultrafast electron dynamics in solar energy conversion}.
\newblock \emph{\bibinfo{journal}{Chem. Rev.}} \textbf{\bibinfo{volume}{117}}, \bibinfo{pages}{10940--11024} (\bibinfo{year}{2017}).

\bibitem{worner2017charge}
\bibinfo{author}{W{\"o}rner, H.~J.} \emph{et~al.}
\newblock \bibinfo{title}{Charge migration and charge transfer in molecular systems}.
\newblock \emph{\bibinfo{journal}{Struct. Dynam.}} \textbf{\bibinfo{volume}{4}}, \bibinfo{pages}{061508} (\bibinfo{year}{2017}).

\bibitem{Scholes2011SolarHarvesting}
\bibinfo{author}{Scholes, G.~D.}, \bibinfo{author}{Fleming, G.~R.}, \bibinfo{author}{Olaya-Castro, A.} \& \bibinfo{author}{van Grondelle, R.}
\newblock \bibinfo{title}{Lessons from nature about solar light harvesting}.
\newblock \emph{\bibinfo{journal}{Nat. Chem.}} \textbf{\bibinfo{volume}{3}}, \bibinfo{pages}{763--774} (\bibinfo{year}{2011}).

\bibitem{Yeh2000ElectronLocalization}
\bibinfo{author}{Yeh, A.~T.}, \bibinfo{author}{Shank, C.~V.} \& \bibinfo{author}{McCusker, J.~K.}
\newblock \bibinfo{title}{Ultrafast electron localization dynamics following photo-induced charge transfer}.
\newblock \emph{\bibinfo{journal}{Science}} \textbf{\bibinfo{volume}{289}}, \bibinfo{pages}{935--938} (\bibinfo{year}{2000}).

\bibitem{Maiuri2018FMOCoherence}
\bibinfo{author}{Maiuri, M.}, \bibinfo{author}{Ostroumov, E.~E.}, \bibinfo{author}{Saer, R.~G.}, \bibinfo{author}{Blankenship, R.~E.} \& \bibinfo{author}{Scholes, G.~D.}
\newblock \bibinfo{title}{Coherent wavepackets in the fenna--matthews--olson complex are robust to excitonic-structure perturbations caused by mutagenesis}.
\newblock \emph{\bibinfo{journal}{Nat. Chem.}} \textbf{\bibinfo{volume}{10}}, \bibinfo{pages}{177--183} (\bibinfo{year}{2018}).

\bibitem{zewail2000femtochemistry}
\bibinfo{author}{Zewail, A.~H.}
\newblock \bibinfo{title}{Femtochemistry: Atomic-scale dynamics of the chemical bond}.
\newblock \emph{\bibinfo{journal}{J. Phys. Chem. A}} \textbf{\bibinfo{volume}{104}}, \bibinfo{pages}{5660--5694} (\bibinfo{year}{2000}).

\bibitem{Blanchet1999VibronicDynamics}
\bibinfo{author}{Blanchet, V.}, \bibinfo{author}{Zgierski, M.~Z.}, \bibinfo{author}{Seideman, T.} \& \bibinfo{author}{Stolow, A.}
\newblock \bibinfo{title}{Discerning vibronic molecular dynamics using time-resolved photoelectron spectroscopy}.
\newblock \emph{\bibinfo{journal}{Nature}} \textbf{\bibinfo{volume}{401}}, \bibinfo{pages}{52--54} (\bibinfo{year}{1999}).

\bibitem{Gessner2006MultidimensionalImaging}
\bibinfo{author}{Ge{\ss}ner, O.} \emph{et~al.}
\newblock \bibinfo{title}{Femtosecond multidimensional imaging of a molecular dissociation}.
\newblock \emph{\bibinfo{journal}{Science}} \textbf{\bibinfo{volume}{311}}, \bibinfo{pages}{219--222} (\bibinfo{year}{2006}).

\bibitem{NugentGlandorf2001Br2Dissociation}
\bibinfo{author}{Nugent-Glandorf, L.} \emph{et~al.}
\newblock \bibinfo{title}{Ultrafast time-resolved soft x-ray photoelectron spectroscopy of dissociating Br\textsubscript{2}}.
\newblock \emph{\bibinfo{journal}{Phys. Rev. Lett.}} \textbf{\bibinfo{volume}{87}}, \bibinfo{pages}{193002} (\bibinfo{year}{2001}).

\bibitem{Dantus1990MolecularVibration}
\bibinfo{author}{Dantus, M.}, \bibinfo{author}{Bowman, R.~M.} \& \bibinfo{author}{Zewail, A.~H.}
\newblock \bibinfo{title}{Femtosecond laser observations of molecular vibration and rotation}.
\newblock \emph{\bibinfo{journal}{Nature}} \textbf{\bibinfo{volume}{343}}, \bibinfo{pages}{737--739} (\bibinfo{year}{1990}).

\bibitem{palacios2020quantum}
\bibinfo{author}{Palacios, A.} \& \bibinfo{author}{Mart{\'\i}n, F.}
\newblock \bibinfo{title}{The quantum chemistry of attosecond molecular science}.
\newblock \emph{\bibinfo{journal}{WIRes. Comput. Mol. Sci.}} \textbf{\bibinfo{volume}{10}}, \bibinfo{pages}{e1430} (\bibinfo{year}{2020}).

\bibitem{lepine2014attosecond}
\bibinfo{author}{L{\'e}pine, F.}, \bibinfo{author}{Ivanov, M.~Y.} \& \bibinfo{author}{Vrakking, M.~J.}
\newblock \bibinfo{title}{Attosecond molecular dynamics: fact or fiction?}
\newblock \emph{\bibinfo{journal}{Nat. Photonics}} \textbf{\bibinfo{volume}{8}}, \bibinfo{pages}{195--204} (\bibinfo{year}{2014}).

\bibitem{corkum2007attosecond}
\bibinfo{author}{Corkum, P.~B.} \& \bibinfo{author}{Krausz, F.}
\newblock \bibinfo{title}{Attosecond science}.
\newblock \emph{\bibinfo{journal}{Nat. Phys.}} \textbf{\bibinfo{volume}{3}}, \bibinfo{pages}{381--387} (\bibinfo{year}{2007}).

\bibitem{goulielmakis2010real}
\bibinfo{author}{Goulielmakis, E.} \emph{et~al.}
\newblock \bibinfo{title}{Real-time observation of valence electron motion}.
\newblock \emph{\bibinfo{journal}{Nature}} \textbf{\bibinfo{volume}{466}}, \bibinfo{pages}{739--743} (\bibinfo{year}{2010}).

\bibitem{calegari2014ultrafast}
\bibinfo{author}{Calegari, F.} \emph{et~al.}
\newblock \bibinfo{title}{Ultrafast electron dynamics in phenylalanine initiated by attosecond pulses}.
\newblock \emph{\bibinfo{journal}{Science}} \textbf{\bibinfo{volume}{346}}, \bibinfo{pages}{336--339} (\bibinfo{year}{2014}).

\bibitem{nisoli2017attosecond}
\bibinfo{author}{Nisoli, M.}, \bibinfo{author}{Decleva, P.}, \bibinfo{author}{Calegari, F.}, \bibinfo{author}{Palacios, A.} \& \bibinfo{author}{Mart{\'\i}n, F.}
\newblock \bibinfo{title}{Attosecond electron dynamics in molecules}.
\newblock \emph{\bibinfo{journal}{Chem. Rev.}} \textbf{\bibinfo{volume}{117}}, \bibinfo{pages}{10760--10825} (\bibinfo{year}{2017}).

\bibitem{okino2015direct}
\bibinfo{author}{Okino, T.} \emph{et~al.}
\newblock \bibinfo{title}{Direct observation of an attosecond electron wave packet in a nitrogen molecule}.
\newblock \emph{\bibinfo{journal}{Sci. Adv.}} \textbf{\bibinfo{volume}{1}}, \bibinfo{pages}{e1500356} (\bibinfo{year}{2015}).

\bibitem{matselyukh2022decoherence}
\bibinfo{author}{Matselyukh, D.~T.}, \bibinfo{author}{Despr{\'e}, V.}, \bibinfo{author}{Golubev, N.~V.}, \bibinfo{author}{Kuleff, A.~I.} \& \bibinfo{author}{W{\"o}rner, H.~J.}
\newblock \bibinfo{title}{Decoherence and revival in attosecond charge migration driven by non-adiabatic dynamics}.
\newblock \emph{\bibinfo{journal}{Nat. Phys.}} \textbf{\bibinfo{volume}{18}}, \bibinfo{pages}{1206--1213} (\bibinfo{year}{2022}).

\bibitem{kowalewski2015catching}
\bibinfo{author}{Kowalewski, M.}, \bibinfo{author}{Bennett, K.}, \bibinfo{author}{Dorfman, K.~E.} \& \bibinfo{author}{Mukamel, S.}
\newblock \bibinfo{title}{Catching conical intersections in the act: Monitoring transient electronic coherences by attosecond stimulated x-ray raman signals}.
\newblock \emph{\bibinfo{journal}{Phys. Rev. Lett.}} \textbf{\bibinfo{volume}{115}}, \bibinfo{pages}{193003} (\bibinfo{year}{2015}).

\bibitem{Vacher2017ElectronDynamics}
\bibinfo{author}{Vacher, M.}, \bibinfo{author}{Bearpark, M.~J.}, \bibinfo{author}{Robb, M.~A.} \& \bibinfo{author}{Malhado, J.~P.}
\newblock \bibinfo{title}{Electron dynamics upon ionization of polyatomic molecules: Coupling to quantum nuclear motion and decoherence}.
\newblock \emph{\bibinfo{journal}{Phys. Rev. Lett.}} \textbf{\bibinfo{volume}{118}}, \bibinfo{pages}{083001} (\bibinfo{year}{2017}).

\bibitem{Despre2018ChargeMigration}
\bibinfo{author}{Despr{\'e}, V.}, \bibinfo{author}{Golubev, N.~V.} \& \bibinfo{author}{Kuleff, A.~I.}
\newblock \bibinfo{title}{Charge migration in propiolic acid: A full quantum dynamical study}.
\newblock \emph{\bibinfo{journal}{Phys. Rev. Lett.}} \textbf{\bibinfo{volume}{121}}, \bibinfo{pages}{203002} (\bibinfo{year}{2018}).

\bibitem{PhysRevLett.126.133002}
\bibinfo{author}{Folorunso, A.~S.} \emph{et~al.}
\newblock \bibinfo{title}{Molecular modes of attosecond charge migration}.
\newblock \emph{\bibinfo{journal}{Phys. Rev. Lett.}} \textbf{\bibinfo{volume}{126}}, \bibinfo{pages}{133002} (\bibinfo{year}{2021}).

\bibitem{Kraus2015}
\bibinfo{author}{Kraus, P.~M.} \emph{et~al.}
\newblock \bibinfo{title}{Measurement and laser control of attosecond charge migration in ionized iodoacetylene}.
\newblock \emph{\bibinfo{journal}{Science}} \textbf{\bibinfo{volume}{350}}, \bibinfo{pages}{790--795} (\bibinfo{year}{2015}).

\bibitem{Schwickert2022}
\bibinfo{author}{Schwickert, D.} \emph{et~al.}
\newblock \bibinfo{title}{Electronic quantum coherence in glycine molecules probed with ultrashort x-ray pulses in real time}.
\newblock \emph{\bibinfo{journal}{Sci. Adv.}} \textbf{\bibinfo{volume}{8}}, \bibinfo{pages}{eabn6848} (\bibinfo{year}{2022}).

\bibitem{Guo2024}
\bibinfo{author}{Guo, Z.} \emph{et~al.}
\newblock \bibinfo{title}{Experimental demonstration of attosecond pump–probe spectroscopy with an x-ray free-electron laser}.
\newblock \emph{\bibinfo{journal}{Nat. Photonics}} \textbf{\bibinfo{volume}{18}}, \bibinfo{pages}{691--697} (\bibinfo{year}{2024}).

\bibitem{kretschmar2024compact}
\bibinfo{author}{Kretschmar, M.} \emph{et~al.}
\newblock \bibinfo{title}{Compact realization of all-attosecond pump-probe spectroscopy}.
\newblock \emph{\bibinfo{journal}{Sci. Adv.}} \textbf{\bibinfo{volume}{10}}, \bibinfo{pages}{eadk9605} (\bibinfo{year}{2024}).

\bibitem{eckle2008attosecond}
\bibinfo{author}{Eckle, P.} \emph{et~al.}
\newblock \bibinfo{title}{Attosecond angular streaking}.
\newblock \emph{\bibinfo{journal}{Nat. Phys.}} \textbf{\bibinfo{volume}{4}}, \bibinfo{pages}{565--570} (\bibinfo{year}{2008}).

\bibitem{Belshaw2012ChargeMigration}
\bibinfo{author}{Belshaw, L.} \emph{et~al.}
\newblock \bibinfo{title}{Observation of ultrafast charge migration in an amino acid}.
\newblock \emph{\bibinfo{journal}{J. Phys. Chem. Lett.}} \textbf{\bibinfo{volume}{3}}, \bibinfo{pages}{3751--3754} (\bibinfo{year}{2012}).

\bibitem{parker2009ultrafast}
\bibinfo{author}{Parker, D.}, \bibinfo{author}{Minns, R.}, \bibinfo{author}{Penfold, T.}, \bibinfo{author}{Worth, G.} \& \bibinfo{author}{Fielding, H.}
\newblock \bibinfo{title}{Ultrafast dynamics of the S1 excited state of benzene}.
\newblock \emph{\bibinfo{journal}{Chem. Phys. Lett.}} \textbf{\bibinfo{volume}{469}}, \bibinfo{pages}{43--47} (\bibinfo{year}{2009}).

\bibitem{PhysRevA.110.033118}
\bibinfo{author}{Kaufman, B.}, \bibinfo{author}{Marquetand, P.}, \bibinfo{author}{Rozgonyi, T.} \& \bibinfo{author}{Weinacht, T.}
\newblock \bibinfo{title}{Long-lived electronic coherences in molecular wave packets probed with pulse-shape spectroscopy}.
\newblock \emph{\bibinfo{journal}{Phys. Rev. A}} \textbf{\bibinfo{volume}{110}}, \bibinfo{pages}{033118} (\bibinfo{year}{2024}).

\bibitem{PhysRevLett.131.263202}
\bibinfo{author}{Kaufman, B.}, \bibinfo{author}{Marquetand, P.}, \bibinfo{author}{Rozgonyi, T.} \& \bibinfo{author}{Weinacht, T.}
\newblock \bibinfo{title}{Long-lived electronic coherences in molecules}.
\newblock \emph{\bibinfo{journal}{Phys. Rev. Lett.}} \textbf{\bibinfo{volume}{131}}, \bibinfo{pages}{263202} (\bibinfo{year}{2023}).

\bibitem{mougol2024direct}
\bibinfo{author}{Mo{\u{g}}ol, G.}, \bibinfo{author}{Kaufman, B.}, \bibinfo{author}{Weinacht, T.}, \bibinfo{author}{Cheng, C.} \& \bibinfo{author}{Ben-Itzhak, I.}
\newblock \bibinfo{title}{Direct observation of entangled electronic-nuclear wave packets}.
\newblock \emph{\bibinfo{journal}{Phys. Rev. Res.}} \textbf{\bibinfo{volume}{6}}, \bibinfo{pages}{L022047} (\bibinfo{year}{2024}).

\bibitem{palacios2014molecular}
\bibinfo{author}{Palacios, A.}, \bibinfo{author}{Gonz{\'a}lez-Castrillo, A.} \& \bibinfo{author}{Mart{\'\i}n, F.}
\newblock \bibinfo{title}{Molecular interferometer to decode attosecond electron--nuclear dynamics}.
\newblock \emph{\bibinfo{journal}{PNAS}} \textbf{\bibinfo{volume}{111}}, \bibinfo{pages}{3973--3978} (\bibinfo{year}{2014}).

\bibitem{gonzalez2020quantum}
\bibinfo{author}{Gonz{\'a}lez-Castrillo, A.}, \bibinfo{author}{Mart{\'\i}n, F.} \& \bibinfo{author}{Palacios, A.}
\newblock \bibinfo{title}{Quantum state holography to reconstruct the molecular wave packet using an attosecond XUV--XUV pump-probe technique}.
\newblock \emph{\bibinfo{journal}{Sci. Rep.}} \textbf{\bibinfo{volume}{10}}, \bibinfo{pages}{12981} (\bibinfo{year}{2020}).

\bibitem{PhysRevLett.128.043201}
\bibinfo{author}{Koll, L.-M.}, \bibinfo{author}{Maikowski, L.}, \bibinfo{author}{Drescher, L.}, \bibinfo{author}{Witting, T.} \& \bibinfo{author}{Vrakking, M. J.~J.}
\newblock \bibinfo{title}{Experimental control of quantum-mechanical entanglement in an attosecond pump-probe experiment}.
\newblock \emph{\bibinfo{journal}{Phys. Rev. Lett.}} \textbf{\bibinfo{volume}{128}}, \bibinfo{pages}{043201} (\bibinfo{year}{2022}).

\bibitem{Vrakking2021AttosecondControl}
\bibinfo{author}{Vrakking, M. J.~J.}
\newblock \bibinfo{title}{Control of attosecond entanglement and coherence}.
\newblock \emph{\bibinfo{journal}{Phys. Rev. Lett.}} \textbf{\bibinfo{volume}{126}}, \bibinfo{pages}{113203} (\bibinfo{year}{2021}).

\bibitem{Koll:22}
\bibinfo{author}{Koll, L.-M.}, \bibinfo{author}{Maikowski, L.}, \bibinfo{author}{Drescher, L.}, \bibinfo{author}{Vrakking, M. J.~J.} \& \bibinfo{author}{Witting, T.}
\newblock \bibinfo{title}{Phase-locking of time-delayed attosecond XUV pulse pairs}.
\newblock \emph{\bibinfo{journal}{Opt. Express}} \textbf{\bibinfo{volume}{30}}, \bibinfo{pages}{7082--7095} (\bibinfo{year}{2022}).

\bibitem{blanchet1997temporal}
\bibinfo{author}{Blanchet, V.}, \bibinfo{author}{Nicole, C.}, \bibinfo{author}{Bouchene, M.-A.} \& \bibinfo{author}{Girard, B.}
\newblock \bibinfo{title}{Temporal coherent control in two-photon transitions: from optical interferences to quantum interferences}.
\newblock \emph{\bibinfo{journal}{Phys. Rev. Lett.}} \textbf{\bibinfo{volume}{78}}, \bibinfo{pages}{2716} (\bibinfo{year}{1997}).

\bibitem{feist2011attosecond}
\bibinfo{author}{Feist, J.} \emph{et~al.}
\newblock \bibinfo{title}{Attosecond two-photon interferometry for doubly excited states of helium}.
\newblock \emph{\bibinfo{journal}{Phys. Rev. Lett.}} \textbf{\bibinfo{volume}{107}}, \bibinfo{pages}{093005} (\bibinfo{year}{2011}).

\bibitem{engel1994two}
\bibinfo{author}{Engel, V.} \& \bibinfo{author}{Metiu, H.}
\newblock \bibinfo{title}{Two-photon wave-packet interferometry}.
\newblock \emph{\bibinfo{journal}{J. Chem. Phys.}} \textbf{\bibinfo{volume}{100}}, \bibinfo{pages}{5448--5458} (\bibinfo{year}{1994}).

\bibitem{blanchet1998temporal}
\bibinfo{author}{Blanchet, V.}, \bibinfo{author}{Bouch{\`e}ne, M.~A.} \& \bibinfo{author}{Girard, B.}
\newblock \bibinfo{title}{Temporal coherent control in the photoionization of Cs$_2$: Theory and experiment}.
\newblock \emph{\bibinfo{journal}{J. Chem. Phys.}} \textbf{\bibinfo{volume}{108}}, \bibinfo{pages}{4862--4876} (\bibinfo{year}{1998}).

\bibitem{ohmori2006real}
\bibinfo{author}{Ohmori, K.} \emph{et~al.}
\newblock \bibinfo{title}{Real-time observation of phase-controlled molecular wave-packet interference}.
\newblock \emph{\bibinfo{journal}{Phys. Rev. Lett.}} \textbf{\bibinfo{volume}{96}}, \bibinfo{pages}{093002} (\bibinfo{year}{2006}).

\bibitem{mudrich2008quantum}
\bibinfo{author}{Mudrich, M.}, \bibinfo{author}{Stienkemeier, F.}, \bibinfo{author}{Droppelmann, G.}, \bibinfo{author}{Claas, P.} \& \bibinfo{author}{Schulz, C.}
\newblock \bibinfo{title}{Quantum interference spectroscopy of rubidium-helium exciplexes formed on helium nanodroplets}.
\newblock \emph{\bibinfo{journal}{Phys. Rev. Lett.}} \textbf{\bibinfo{volume}{100}}, \bibinfo{pages}{023401} (\bibinfo{year}{2008}).

\bibitem{morrigan2023ultrafast}
\bibinfo{author}{Morrigan, L.} \emph{et~al.}
\newblock \bibinfo{title}{Ultrafast molecular frame quantum tomography}.
\newblock \emph{\bibinfo{journal}{Phys. Rev. Lett.}} \textbf{\bibinfo{volume}{131}}, \bibinfo{pages}{193001} (\bibinfo{year}{2023}).

\bibitem{baumgartner2022ultrafast}
\bibinfo{author}{Baumg{\"a}rtner, K.} \emph{et~al.}
\newblock \bibinfo{title}{Ultrafast orbital tomography of a pentacene film using time-resolved momentum microscopy at a FEL}.
\newblock \emph{\bibinfo{journal}{Nat. Commun.}} \textbf{\bibinfo{volume}{13}}, \bibinfo{pages}{2741} (\bibinfo{year}{2022}).

\bibitem{born1996dynamical}
\bibinfo{author}{Born, M.} \& \bibinfo{author}{Huang, K.}
\newblock \emph{\bibinfo{title}{Dynamical Theory of Crystal Lattices}}  (\bibinfo{publisher}{Oxford University Press}, \bibinfo{year}{1996}).

\bibitem{stolow2004femtosecond}
\bibinfo{author}{Stolow, A.}, \bibinfo{author}{Bragg, A.~E.} \& \bibinfo{author}{Neumark, D.~M.}
\newblock \bibinfo{title}{Femtosecond time-resolved photoelectron spectroscopy}.
\newblock \emph{\bibinfo{journal}{Chem. Rev.}} \textbf{\bibinfo{volume}{104}}, \bibinfo{pages}{1719--1758} (\bibinfo{year}{2004}).

\bibitem{gerber1991Na2}
\bibinfo{author}{Baumert, T.}, \bibinfo{author}{Grosser, M.}, \bibinfo{author}{Thalweiser, R.} \& \bibinfo{author}{Gerber, G.}
\newblock \bibinfo{title}{Femtosecond time-resolved molecular multiphoton ionization: The ${\mathrm{Na}}_{2}$ system}.
\newblock \emph{\bibinfo{journal}{Phys. Rev. Lett.}} \textbf{\bibinfo{volume}{67}}, \bibinfo{pages}{3753--3756} (\bibinfo{year}{1991}).

\bibitem{shim2009turn}
\bibinfo{author}{Shim, S.-H.} \& \bibinfo{author}{Zanni, M.~T.}
\newblock \bibinfo{title}{How to turn your pump--probe instrument into a multidimensional spectrometer: 2d IR and Vis spectroscopies via pulse shaping}.
\newblock \emph{\bibinfo{journal}{Phys. Chem. Chem. Phys.}} \textbf{\bibinfo{volume}{11}}, \bibinfo{pages}{748--761} (\bibinfo{year}{2009}).

\bibitem{tian2003femtosecond}
\bibinfo{author}{Tian, P.}, \bibinfo{author}{Keusters, D.}, \bibinfo{author}{Suzaki, Y.} \& \bibinfo{author}{Warren, W.~S.}
\newblock \bibinfo{title}{Femtosecond phase-coherent two-dimensional spectroscopy}.
\newblock \emph{\bibinfo{journal}{Science}} \textbf{\bibinfo{volume}{300}}, \bibinfo{pages}{1553--1555} (\bibinfo{year}{2003}).

\bibitem{pearson2007shaped}
\bibinfo{author}{Pearson, B.~J.} \& \bibinfo{author}{Weinacht, T.~C.}
\newblock \bibinfo{title}{Shaped ultrafast laser pulses in the deep ultraviolet}.
\newblock \emph{\bibinfo{journal}{Opt. Express}} \textbf{\bibinfo{volume}{15}}, \bibinfo{pages}{4385--4388} (\bibinfo{year}{2007}).

\bibitem{codere2024high}
\bibinfo{author}{Codere, J.} \emph{et~al.}
\newblock \bibinfo{title}{High repetition-rate pulse shaping of a spectrally broadened Yb femtosecond laser}.
\newblock \emph{\bibinfo{journal}{Opt. Contin.}} \textbf{\bibinfo{volume}{3}}, \bibinfo{pages}{785--794} (\bibinfo{year}{2024}).

\bibitem{catanese2021acousto}
\bibinfo{author}{Catanese, A.} \emph{et~al.}
\newblock \bibinfo{title}{Acousto-optic modulator pulse-shaper compression of octave-spanning pulses from a stretched hollow-core fiber}.
\newblock \emph{\bibinfo{journal}{OSA Contin.}} \textbf{\bibinfo{volume}{4}}, \bibinfo{pages}{3176--3183} (\bibinfo{year}{2021}).

\bibitem{codere2025spectral}
\bibinfo{author}{Codere, J.} \emph{et~al.}
\newblock \bibinfo{title}{Spectral broadening and pulse shaping in the deep ultraviolet}.
\newblock \emph{\bibinfo{journal}{Opt. Lett.}} \textbf{\bibinfo{volume}{51}}, \bibinfo{pages}{269--272} (\bibinfo{year}{2025}).

\bibitem{lee2024few}
\bibinfo{author}{Lee, J.~P.} \emph{et~al.}
\newblock \bibinfo{title}{Few-femtosecond soft x-ray transient absorption spectroscopy with tuneable DUV-Vis pump pulses}.
\newblock \emph{\bibinfo{journal}{Optica}} \textbf{\bibinfo{volume}{11}}, \bibinfo{pages}{1320--1323} (\bibinfo{year}{2024}).

\bibitem{huppert2015attosecond}
\bibinfo{author}{Huppert, M.}, \bibinfo{author}{Jordan, I.} \& \bibinfo{author}{W{\"o}rner, H.~J.}
\newblock \bibinfo{title}{Attosecond beamline with actively stabilized and spatially separated beam paths}.
\newblock \emph{\bibinfo{journal}{Rev. Sci. Instrum.}} \textbf{\bibinfo{volume}{86}}, \bibinfo{pages}{123106} (\bibinfo{year}{2015}).

\bibitem{borrego2016two}
\bibinfo{author}{Borrego-Varillas, R.} \emph{et~al.}
\newblock \bibinfo{title}{Two-dimensional electronic spectroscopy in the ultraviolet by a birefringent delay line}.
\newblock \emph{\bibinfo{journal}{Opt. Express}} \textbf{\bibinfo{volume}{24}}, \bibinfo{pages}{28491--28499} (\bibinfo{year}{2016}).

\bibitem{herzberg1966electronic}
\bibinfo{author}{Herzberg, G.}
\newblock \emph{\bibinfo{title}{Molecular Spectra and Molecular Structure. Volume III: Electronic Spectra and Electronic Structure of Polyatomic Molecules}} (\bibinfo{publisher}{Van Nostrand Reinhold}, \bibinfo{address}{New York}, \bibinfo{year}{1966}).

\bibitem{wilson1980molecular}
\bibinfo{author}{Wilson, E.~B.}, \bibinfo{author}{Decius, J.~C.} \& \bibinfo{author}{Cross, P.~C.}
\newblock \emph{\bibinfo{title}{Molecular Vibrations: The Theory of Infrared and Raman Vibrational Spectra}}  (\bibinfo{publisher}{Courier Corporation}, \bibinfo{year}{1980}).

\bibitem{weinacht2018time}
\bibinfo{author}{Weinacht, T.} \& \bibinfo{author}{Pearson, B.~J.}
\newblock \emph{\bibinfo{title}{Time-Resolved Spectroscopy: An Experimental Perspective}}  (\bibinfo{publisher}{CRC Press}, \bibinfo{year}{2018}).

\bibitem{hamm2011concepts}
\bibinfo{author}{Hamm, P.} \& \bibinfo{author}{Zanni, M.}
\newblock \emph{\bibinfo{title}{Concepts and Methods of 2D Infrared Spectroscopy}}  (\bibinfo{publisher}{Cambridge University Press}, \bibinfo{year}{2011}).

\bibitem{Fuller20152DES}
\bibinfo{author}{Fuller, F.~D.} \& \bibinfo{author}{Ogilvie, J.~P.}
\newblock \bibinfo{title}{Experimental implementations of two-dimensional fourier transform electronic spectroscopy}.
\newblock \emph{\bibinfo{journal}{Ann. Rev. Phys. Chem.}} \textbf{\bibinfo{volume}{66}}, \bibinfo{pages}{667--690} (\bibinfo{year}{2015}).

\bibitem{potts1972photoelectron}
\bibinfo{author}{Potts, A.}, \bibinfo{author}{Price, W.}, \bibinfo{author}{Streets, D.} \& \bibinfo{author}{Williams, T.}
\newblock \bibinfo{title}{Photoelectron spectra of benzene and some fluorobenzenes}.
\newblock \emph{\bibinfo{journal}{Faraday Discuss.}} \textbf{\bibinfo{volume}{54}}, \bibinfo{pages}{168--181} (\bibinfo{year}{1972}).

\bibitem{rozgonyi2000application}
\bibinfo{author}{Rozgonyi, T.}, \bibinfo{author}{Gla{\ss}, A.} \& \bibinfo{author}{Feurer, T.}
\newblock \bibinfo{title}{Application of nonreflecting boundary condition for numerical simulation of molecular photoionization dynamics}.
\newblock \emph{\bibinfo{journal}{J. Appl. Phys.}} \textbf{\bibinfo{volume}{88}}, \bibinfo{pages}{2936--2942} (\bibinfo{year}{2000}).

\bibitem{pantos1978extinction}
\bibinfo{author}{Pantos, E.}, \bibinfo{author}{Philis, J.} \& \bibinfo{author}{Bolovinos, A.}
\newblock \bibinfo{title}{The extinction coefficient of benzene vapor in the region 4.6 to 36 eV}.
\newblock \emph{\bibinfo{journal}{J. Mol. Spectrosc.}} \textbf{\bibinfo{volume}{72}}, \bibinfo{pages}{36--43} (\bibinfo{year}{1978}).

\bibitem{Meyer1990}
\bibinfo{author}{Meyer, H.-D.}, \bibinfo{author}{Manthe, U.} \& \bibinfo{author}{Cederbaum, L.~S.}
\newblock \bibinfo{title}{The multi-configurational time-dependent Hartree approach}.
\newblock \emph{\bibinfo{journal}{Chem. Phys. Lett.}} \textbf{\bibinfo{volume}{165}}, \bibinfo{pages}{73--78} (\bibinfo{year}{1990}).

\bibitem{brixner:2D_dots}
\bibinfo{author}{Mueller, S.}, \bibinfo{author}{L{\"u}ttig, J.}, \bibinfo{author}{Brenneis, L.}, \bibinfo{author}{Oron, D.} \& \bibinfo{author}{Brixner, T.}
\newblock \bibinfo{title}{Observing Multiexciton Correlations in Colloidal Semiconductor Quantum Dots via Multiple-Quantum Two-Dimensional Fluorescence Spectroscopy}.
\newblock \emph{\bibinfo{journal}{ACS Nano}} \textbf{\bibinfo{volume}{15}}, \bibinfo{pages}{4647--4657} (\bibinfo{year}{2021}).
\bibitem{data}
\bibinfo{author}{Ngo, L. T.-H.} \emph{et~al.}
\newblock \bibinfo{title}{Source Data for "Molecular Attoscope: Pulse Shape Spectroscopy of Electronic Coherence" https://doi.org/10.6084/m9.figshare.31224808} (\bibinfo{year}{2026}).

\end{thebibliography}
\end{document}